\documentclass[letterpaper, 10 pt, article]{article}  

  \let\labelindent\relax
\usepackage{amsmath} 
\usepackage{amssymb}  

\usepackage{tabulary}
\usepackage[english]{babel}
\usepackage{blindtext}
\usepackage[font=footnotesize]{caption}
\usepackage[makeroom]{cancel}
\usepackage{amsfonts}
\usepackage{pgfplots}
\usepackage{amsthm}  
\usepackage{amssymb}
\usepackage{graphicx}
\usepackage{pgf,tikz}
\pgfplotsset{ 
  compat=newest, 
   legend style =
  {font=\footnotesize },
  label style = {font=\footnotesize},
every tick label/.append style={font=\footnotesize}
  }
\usepackage{tabularx}
\usepackage{cite}
\usepgfplotslibrary{fillbetween}
  \usepgfplotslibrary{ternary}
\usetikzlibrary{shapes.multipart}
\usepackage{enumitem} 
\usepackage{bbm} 
\usepackage{amsmath}
\usepackage{color}
\usepackage{graphicx}
\usepackage{amsfonts}
\usepackage{dsfont}
\usepackage{tikz}
\usepackage{caption}
\usepackage{subcaption}
\usepackage{amsthm}
\usepackage{nicematrix}

\usepackage[strict]{changepage}

\newcommand{\vect}[1]{\boldsymbol{#1}}

\newcommand{\mc}{\mathcal}
\newcommand{\be}{\begin{equation}}
\newcommand{\ee}{\end{equation}}
\allowdisplaybreaks

\newtheorem{remark}{Remark}
\newtheorem{theorem}{Theorem}
\newtheorem{lemma}{Lemma}
\newtheorem{assumption}{Assumption}

\newtheorem{definition}{Definition}
\newtheorem{corollary}{Corollary}
\newtheorem{proposition}{Proposition}

\newtheorem{example}{Example}
\newtheorem{usecase}{Use Case}

\definecolor{mygreen}{RGB}{72,160,66}
\definecolor{myblue}{RGB}{20,155,204}
\definecolor{myyellow}{RGB}{190,170,0}
\definecolor{myviolet}{RGB}{195,8,255}
\definecolor{myred}{RGB}{204,0,0}
\definecolor{myblue}{RGB}{0,66,255}
\definecolor{mycolor1}{rgb}{0.00000,0.44700,0.74100}%
\definecolor{mycolor2}{rgb}{0.85000,0.32500,0.09800}%
\definecolor{mycolor3}{rgb}{0.92900,0.69400,0.12500}%
\definecolor{mycolor4}{rgb}{0.49400,0.18400,0.55600}%
\definecolor{mycolor5}{rgb}{0.46600,0.67400,0.18800}%
\definecolor{mycolor6}{rgb}{0.30100,0.74500,0.93300}%
\definecolor{mycolor7}{rgb}{0.63500,0.07800,0.18400}%

\title{\Large\bf FJ-MM: The Friedkin-Johnsen Opinion Dynamics Model\\ with Memory and Higher-Order Neighbors}
\author{Roberta Raineri$^{a,b}$, Lorenzo Zino$^{a}$ and Anton Proskurnikov$^{a}$
\thanks{$^a$ Department of Electronics and Telecommunications, Politecnico di Torino, Turin, Italy; $^b$ Department of Mathematical Sciences ``G.L. Lagrange'', Politecnico di Torino, Turin, Italy. Email: \texttt{\{roberta.raineri,lorenzo.zino\}@polito.it, anton.p.1982@ieee.org}. A shortened version of this preprint will appear in the proceedings of the European Control Conference 2025.}
}

\begin{document}

\maketitle
\thispagestyle{empty}

\begin{abstract}
The Friedkin-Johnsen (FJ) model has been extensively explored and validated, spanning applications in social science, systems and control, game theory, and algorithmic research. In this paper, we introduce an advanced generalization of the FJ model, termed FJ-MM which incorporates both memory effects and multi-hop (higher-order neighbor) influence. 
This formulation allows agents to naturally incorporate both current and previous opinions at each iteration stage.
Our numerical results demonstrate that incorporating memory and multi-hop influence significantly reshapes the opinion landscape; for example, the final opinion profile can exhibit reduced polarization.
We analyze the stability and equilibrium properties of the FJ-MM model, showing that these properties can be reduced to those of a \emph{comparison} model--namely, the standard FJ model with a modified influence matrix. This reduction enables us to leverage established stability results from FJ dynamics. Additionally, we examine the convergence rate of the FJ-MM model and demonstrate that, as can be expected, the time lags introduced by memory and higher-order neighbor influences result in slower convergence.
\end{abstract}

\section{Introduction}
Agent-based opinion dynamics modeling is a rapidly advancing field that attracts researchers from social sciences, economics, physics, engineering, and beyond; see~\cite{Friedkin:2015,mastroeni2019agent,proskurnikov2018tutorial,Grabisch2020,Anderson2019} for comprehensive reviews of its history and recent developments. The most studied models in engineering and mathematical literature rely on \emph{iterative opinion averaging} as the key driving force of opinion formation, a mechanism originating from pioneering work on social power~\cite{french1956formal} and studies on rational consensus~\cite{DeGroot74,Lehrer1976}. Recent experiments confirm a tendency toward opinion averaging in small groups~\cite{FRIEDKINPROBULLO:2021},
medium-size groups~\cite{Takcs2016}, and large-scale online communities~\cite{kozitsin2022}.
The central element of an averaging-based model is a weighted digraph of social influence, which can be static or co-evolve with opinions. 
This digraph depicts social ties among individuals and quantifies the weights each agent assigns to those they are connected with.
Individuals (nodes in the graph) update their opinions by taking the weighted average of the opinions of adjacent nodes, with updates occurring simultaneously or asynchronously~\cite{proskurnikov2018tutorial}.

\subsection*{The Friedkin-Johnsen (FJ) model}

The Friedkin-Johnsen (FJ) model~\cite{FJ99} is a seminal and extensively studied model of opinion formation, naturally extending the French-DeGroot iterative averaging dynamics. In addition to the weighted influence digraph, the FJ model assigns each agent a constant \emph{innate} opinion, factored into each opinion update iteration. Originally defined as the agent's initial opinion~\cite{FJ99}, the innate opinion can also be shaped by the agent's prejudices or some other sources of information, such as social media.
The strength of an agent's ``anchorage'' to their innate opinion is regulated by an additional parameter, interpreted as the agent's susceptibility to social influence~\cite{Friedkin:2015}; some studies interpret this constant as a measure of conformity under group pressure~\cite{semonsen2019pressure}. While remaining linear, the FJ model can result in diverse distributions of final opinions, ranging from consensus to multimodal polarized states~\cite{Friedkin:2015}. 

In recent years, the FJ model has been studied from systems and control~\cite{Frasca2013,GHADERI20143209,ravazzi2021learning}, game-theoretic~\cite{Bindel2015,ferraioli2019pressure}, and algorithmic~\cite{Abebe2018,Xu2021,Neumann2024} perspectives; a number of experiments have been conducted to validate it~\cite{FJ99,FriedkinPNAS,FriedProsk2019,FRIEDKINPROBULLO:2021,Friedkin2016}. The FJ model has been extended to describe dynamics of multidimensional opinions on interrelated topics~\cite{Parsegov_CDC2015} and the dynamics of expressed versus private opinions~\cite{ye2019influence}.

\subsection*{The Model Under Study}

In this paper, we analyze a generalization of the FJ model with hereditary effects, where at each opinion update, an agent is influenced not only by current opinions (their own and others') but also by opinions from previous steps. Note that the primary motivation for introducing this model is \textbf{not} to account for communication delays, which are frequently addressed in the opinion dynamics literature~\cite{LiuBasar2016,Kim2017,ChoiPignotti2021} and studies on multi-agent systems~\cite{Blondel2005}.  Rather, the motivation stems from intertwined effects of \emph{higher-order neighbors} in the influence graph, or the multi-hop\footnote{The term ``multi-hop'' refers to indirect (or long-range) influence or communication involving multiple intermediary nodes between the source and target in a network. Multi-hop protocols, utilizing information from higher-order neighbors, have been proposed to accelerate consensus seeking~\cite{Jin2006_MultiHop,Chen2015SecondOrder}.}, as explored in recent works~\cite{Zhang2020_FJwithHigherOrder,Zhang2024}, and individual memory. 
In view of this, we refer to the model in question as the \textbf{FJ-MM}: the FJ model with Memory and Multi-hop influence.

The original FJ model entails \emph{indirect} influence through walks in the influence graph~\cite{Friedkin:2015}. For example, if agent $i$ trusts agent $j$, and agent $j$'s opinions depend on agent $k$, then $k$ indirectly influences $i$ without direct opinion sharing. In~\cite{Zhang2020_FJwithHigherOrder,Zhang2024}, the FJ model is extended to assume that agents average opinions from \emph{long-range} connections (via walks of a given length) along with those of adjacent nodes in each update. Though the exact mechanism of this multi-hop influence is unspecified in~\cite{Zhang2020_FJwithHigherOrder,Zhang2024}, a plausible explanation is that agents share not only their own opinions but also information from their nearest neighbors, thereby disseminating those opinions to individuals who would otherwise lack access to them. An additional mechanism is observational learning, where information about an agent's neighbors' opinions can be inferred by observing the agent’s actions that are influenced by those opinions~\cite{Varma2020}. Both explanations imply that secondary neighbor influence involves a time lag, which is neglected in~\cite{Zhang2020_FJwithHigherOrder,Zhang2024}. For instance, if opinions from agent $k$ reach agent $i$ via the path $i \to j \to k$ at timestep $t$, then agent $j$ must have received information about $k$ at timestep $t-1$ or earlier to relay (or allow inference of) it at time $t$. Essentially, agent $i$ at time $t$ relies on agent $j$'s \emph{memory} of past interactions.

On the other hand, memory effects extend beyond multi-hop opinion propagation. Many social media platforms (e.g., Facebook) provide personalized prompts for users to revisit and reshare historical content, such as past posts and images~\cite{Jacobsen2021}, and may also direct users back to old discussion threads. It is therefore plausible to assume that agents assign positive influence weights not only to current opinions but also to their own and their neighbors' past opinions. Whereas the effects of memory have been explored in consensus-type algorithms for iterative averaging (see, e.g.,~\cite{LiuCaoMorse2013} and references therein), models examining memory’s impact on opinion formation are scarce~\cite{LiuChai2023,Sun2024LSTM,Jedr2018,Becchetti2023}, and to the best of our knowledge, none are based on the FJ linear mechanism.

\subsection*{Structure and Contributions of the Paper}

The remainder of the paper is organized as follows. 
Section~\ref{sec.setup} introduces the FJ-MM model and related concepts, along with preliminaries from graph theory.
The stability analysis of the FJ-MM model is presented in Section~\ref{sec.stab}.
Using results from positive systems theory, we show that the \emph{stability and equilibrium properties} of the FJ-MM reduce to those of the \emph{comparison} system, being the FJ model with a modified matrix. This section also illustrates the impact of memory and multi-hop influence on opinion formation outcomes. In Section~\ref{sec.rate}, we present some findings on convergence rate of the FJ-MM model. Unlike stability, this rate (the Perron-Frobenius eigenvalue) is not determined by the comparison model and depends on the graph topology. Section~\ref{sec.concl} concludes the paper.

\section{Model Definition and Framework}\label{sec.setup}

In this section, we introduce the FJ-MM model and related concepts that will be used in subsequent sections.

\subsection{Preliminaries and Notation} 

We use $\mathbb{R}$, $\mathbb{R}^n$, and $\mathbb{R}^{m\times n}$ to denote, respectively, the sets of real numbers, real $n$-dimensional (column) vectors, and real $m \times n$ matrices. The all-$0$ and all-$1$ column vectors are denoted by 
$\vect{0}_n$ and $\vect{1}_n \in \mathbb{R}^n$ respectively, and the identity $n \times n$ matrix is denoted by $I_n$, with dimensions omitted when clear from context. 
Throughout the text, capital letters denote matrices, and their entries are denoted by lowercase letters—for example, $W = (w_{ij})$.
For vectors and matrices, the relations $\geq,>$ and $\leq,<$ are understood entry-wise. 
Given a vector $v \in \mathbb{R}^n$, the symbol $[v]$ denotes the diagonal matrix $D \in \mathbb{R}^{n \times n}$ with diagonal entries $d_{ii} = v_i$. Given an arbitrary matrix $M$, we denote by $\text{diag}(M)$ the diagonal matrix whose diagonal entries are the same as those of $M$.

A matrix $M$ is Schur stable if $\rho(M) < 1$, where $\rho(M)$ is the spectral radius, i.e., the largest modulus of $M$’s eigenvalues. The well-known Perron–Frobenius theorem~\cite{Matrix} states that for any nonnegative matrix $M \geq 0$, the spectral radius $\rho(M)$ is a (real) eigenvalue of $M$. A matrix $M\geq 0$ is (row) stochastic if $M\vect{1} = \vect{1}$, and (row) substochastic if $M\vect{1} \leq \vect{1}$. 
Using the Ger{\u s}gorin disk theorem~\cite{Matrix}, it is easy to verify that $\rho(M) \leq 1$ for all substochastic matrices, and $\rho(M) = 1$ if $M$ is stochastic.

A (weighted directed) graph is defined by the triple $\mc G[W] = (\mc V, \mc E, W)$, where the set of nodes $\mc V = \{1, \dots, n\}$ represents the individuals (agents), the set of directed edges $\mc E \subseteq \mc V \times \mc V$ indicates the presence of social influence, and $W = (w_{ij}) \geq 0$ is the weighted $n\times n$ adjacency matrix such that $w_{ij} > 0$ if and only if $(i,j) \in \mc E$. Additionally,  a binary adjacency matrix $B\in\{0,1\}^{n\times n}$ can also be introduced with $b_{ij}=1$ if and only if $w_{ij}>0$. For two matrices $W^1,W^2\geq 0$, the graph $\mc G[W^1+W^2]$ corresponds to the union of two graphs $\mc G[W^1]$ and $\mc G[W^2]$.

\subsection{The FJ and FJ-MM Models}

Henceforth, we consider a social group where each agent $i$ is characterized at each discrete timestep $t$ by a scalar state $x_i(t) \in \mathbb{R}$, interpreted as their opinion on a particular topic. The overall network state at time $t$ is represented by the vector $x(t) = [x_1(t), \dots, x_n(t)]^\top\in\mathbb{R}^n$.  

The classical FJ model~\cite{FJ99,Friedkin:2015} is defined by three components: the stochastic \emph{influence matrix} $W = (w_{ij})\in\mathbb{R}^{n\times n}$, the diagonal \emph{susceptibility matrix} $\Lambda\in\mathbb{R}^{n\times n}$ with entries $\lambda_{ii}\in[0,1]$, and the vector of \emph{innate opinions} $s\in\mathbb{R}^{n}$. The model assumes that opinions evolve according to the following update rule: 
\begin{equation}\label{eq:fj-classic}
x_i(t+1)=\lambda_{ii}\bar x_{i}(t)+(1-\lambda_{ii})s_i,\;\bar x_{i}(t):=\sum_{j=1}^{n}w_{ij}x_j(t).
\end{equation}
In other words, at each step of the opinion update, an agent's new opinion is determined by their innate opinion $s_i$ and the weighted average of their own and others' opinions, $\bar{x}_i(t)$. The weight $w_{ij} > 0$, assigned by agent $i$ to agent $j$, reflects $i$'s appraisal of $j$—such as recognition of expertise or trust. The coefficient $\lambda_{ii}$ represents an agent's openness to assimilating others’ opinions, or their \emph{susceptibility} to social influence. An agent with $\lambda_{ii} = 0$ is considered 'totally stubborn', fully anchored to their innate opinion $s_i$, while $\lambda_{ii} = 1$ corresponds to the classical French–DeGroot model, where opinions are updated solely through iterative averaging. In compact matrix form, the state vector $x(t)$ evolves according to the following dynamics:
\be \label{eq:dynamical system}
x(t+1)=\Lambda W x(t)+(I-\Lambda)s,\quad\forall t=0,1,\dots
\ee
The properties of the FJ model have been studied in~\cite{Bindel2015,GHADERI20143209,Parsegov_CDC2015,Friedkin:2015,ProTempoCaoFriedkin2017}, to mention a few.


In this paper, we focus on an extended version of the FJ model, where the average of current opinions from~\eqref{eq:fj-classic}, $\bar{x}_i(t)$, computed by each agent $i$, is expanded to incorporate some opinions from previous steps as follows:
\begin{equation}\label{eq:fj-mm}
\bar x_i(t)=\sum\nolimits_{j=1}^n\sum\nolimits_{\ell=1}^{L}w_{ij}^{(\ell)}x_j(t-\ell+1),
\end{equation}
which leads to the matrix equation
\be \label{eq:FJ_HigherOrder}
x(t+1) = \Lambda \sum\nolimits_{\ell=1}^{L}  W^{(\ell)}x(t-\ell+1) + (I-\Lambda)s.
\ee
Here, $L \geq 1$ represents the depth of memory (with $L=1$ corresponding to the original FJ model~\eqref{eq:fj-classic}), while the convex combination mechanism is preserved: $w_{ij}^{(\ell)} \geq 0$ and $\sum_{\ell=1}^{L} \sum_{j=1}^n w_{ij}^{(\ell)} = 1$. Although this extension can accommodate communication delays, it is not motivated by them; rather, it aims to capture the effects of memory and multi-hop influence, thereby justifying the acronym \textbf{FJ-MM}.

\begin{definition}[FJ-MM]
The system~\eqref{eq:FJ_HigherOrder}, defined by the diagonal matrix $\mathbf{0}\leq\Lambda\leq I_n$ and nonnegative matrices $W^{(\ell)}$, $\ell=1,\ldots,L$,
whose sum $W^{(1)}+\ldots+W^{(L)}$ is a \emph{stochastic} matrix, is referred to as the FJ-MM model.
\end{definition}

Note that the classical FJ model includes the French–DeGroot dynamics as a special case in which all agents are maximally susceptible, i.e., $\Lambda = I_n$. In the FJ-MM model, this corresponds to the French–DeGroot dynamics with memory—a model that has been studied in the context of delay robustness in consensus algorithms~\cite{Blondel2005}. Remarkably, delay can even facilitate consensus in the presence of periodic communication graphs~\cite{ChenLu2017}, where the undelayed DeGroot model is known to oscillate. In this paper, we are primarily interested in the generic case where the FJ-MM dynamics is asymptotically stable, which is only possible when $\lambda_{ii} < 1$ for at least one agent $i$.

We now present two remarks regarding the choice of initial condition and the Lyapunov stability of the FJ-MM.

\begin{remark}[Initial Condition vs. Innate Opinions]
The initial condition of the FJ-MM model is given by the sequence $x(-L+1), \dots, x(0)$. In the original FJ model ($L=1$), it is often assumed that $x(0) = s$, as the innate opinions, according to~\cite{FJ99}, retain information about the agents' past experiences and thus serve a role similar to the initial state vector $x(0)$. Following this logic, a natural choice for the initial condition is $x(-L+1) = \ldots = x(0) = s$. This choice, however, is not crucial as we are primarily interested in the asymptotic stability of the FJ-MM system, which implies that initial conditions are forgotten at an exponential rate.
\end{remark}
\begin{remark}[Nested Convex Hulls]\label{rem.convex}
It is known~\cite{FriedProsk2019} that the FJ model with $s=x(0)$ is featured by the nested convex hull property: the convex hull spanned by the opinions $x_i(t)$ is non-expanding; in particular, the opinions never leave the convex hull of the initial opinions, being an ``implicit'' decision space for the agents. A more general property for the system~\eqref{eq:FJ_HigherOrder} can be proved: the sequences
\[
\begin{gathered}
m(t):=\min\nolimits_{\ell=1,\dots,L}\min\nolimits_i\{x_i(t-\ell+1),s_i\},\\
M(t):=\max\nolimits_{\ell=1,\dots,L}\max\nolimits_i\{x_i(t-\ell+1),s_i\}.    
\end{gathered}
\]
are monotone: $m(t+1)\geq m(t)$ and $M(t+1)\leq M(t)$.
\end{remark}

Using Remark~\ref{rem.convex} and induction on $t$, the following proposition is immediate, entailing that the FJ-MM system is (marginally) Lyapunov stable and has bounded solutions.
\begin{proposition}
    For every solution $x(t)$ it holds that $x_i(t)\in[m(0),M(0)]$ for all $i\in\mathcal V$ and all $t\geq 0$. Hence, the dynamics of the FJ-MM system~\eqref{eq:FJ_HigherOrder} is marginally Lyapunov stable (all solutions are bounded).
\end{proposition}



\subsection{Main Use Cases}

We illustrate the flexibility of the multiple influence weight matrices $W^{(\ell)}$ by considering several scenarios (Use Cases 1–4) that generalize the standard Friedkin–Johnsen social influence networks~\cite{Friedkin:2015}. We adopt the following assumption for brevity and simplicity in the remainder of this paper.
\begin{assumption}[One-Step Memory]\label{a:2}
    The FJ-MM model~\eqref{eq:FJ_HigherOrder} has the depth of memory $L=2$, being thus
\be \label{eq:dynamical system with delay}
    x(t+1)= \Lambda\left(W^{(1)}x(t)+W^{(2)}x(t-1)\right)+(I-\Lambda)s.
\ee
\end{assumption}

In all use cases we consider, the influence matrices are given by
\begin{equation}\label{eq:w-matrices} W^{(1)} = (I - [\beta]) W, \quad W^{(2)} = [\beta] \tilde{W}, \end{equation}
where $W$ and $\tilde{W}$ are stochastic matrices, and $\beta \in [0,1]^n$ is some vector. In other words, the weighted average of the neighbors' opinions in~\eqref{eq:fj-mm} at each time $t$ can be expressed as:
\[
\bar x_i(t)=(1-\beta_i)\sum_j w_{ij}x_j(t)+\beta_i\sum_j\tilde w_{ij}x_j(t-1),\;\;\text{where $\sum_j w_{ij}=\sum_j\tilde w_{ij}=1$}.
\]
The parameter $\beta_i \in [0,1]$ admits a simple interpretation: it represents the total influence weight that agent $i$ allocates to the \emph{past} opinions of herself and others, i.e., $\beta_i = \sum_j (\beta_i \tilde{w}_{ij})$, while the remaining weight, $1 - \beta_i$, is distributed across the \emph{current} opinions. As in the original FJ model (corresponding to $\beta=\mathbf{0}$), the weights $w_{ij}$ and $\tilde{w}_{ij}$ reflect the level of trust that agent $i$ places in the current and past opinions of agent $j$, respectively. However, as already noted, the mechanisms by which agent $i$ receives the current and past opinions of agent $j$ can differ fundamentally: while current opinions are directly communicated by other individuals, past opinions may be accessible only through ``rumors'' spread by them or may rely on their memory. For these reasons, $W$ and $\tilde{W}$ are not only distinct matrices, but may also correspond to entirely different graphs.

\begin{usecase}[Secondary Neighbors]\label{ex:2hop}
Our first use case is inspired by the model in~\cite{Zhang2020_FJwithHigherOrder}, where agents receive opinions from both direct and secondary neighbors in the influence graph $\mathcal{G}[W]$, defined by stochastic matrix $W$, whereas
$\tilde W=W^2$ in~\eqref{eq:w-matrices}. If agent $i$ accesses the opinion of agent $k$ through an intermediary $j$, the weight assigned to $k$'s opinion is proportional to the product $w_{ij} w_{jk}$. Considering all possible two-step walks from $i$ to $k$, the total weight is proportional to the sum of these contributions -- that is, the $(i,k)$ entry of the (weighted) walk matrix $W^2$.

Unlike the model in~\cite{Zhang2020_FJwithHigherOrder}, which assumes immediate availability of secondary neighbors' opinions, we assume that an opinion reaching agent $i$ via the path $i \to j \to k$ at time $t$ is $x_k(t-1)$ rather than $x_k(t)$, as it reflects agent $j$’s memory of an earlier interaction with $k$. 
    \end{usecase}
    \begin{usecase}[Secondary Neighbors, Alternative Weighting]\label{ex:gossip}
 The previous scenario assumes that, when weighting the opinion along the walk $i \to j \to k$, agent $i$ knows the weight $w_{jk}$ assigned by $j$ to $k$ and uses it to compute $\tilde w_{ik}$. Alternatively, if $w_{jk}$ is private to $j$, it is natural to assign equal weights to all secondary neighbors accessible through $j$, with their sum proportional to $w_{ij}$—the more trust $i$ places in $j$, the greater the weight assigned to opinions relayed by $j$. For multiple walks between $i$ and $k$, it is natural to assume that their total weight $\tilde w_{ik}$ is proportional to $\sum_{j=1}^n w_{ij} b_{jk}$, where $b_{jk}$ is the entry of the binary adjacency matrix for $\mc G[W]$. Hence, it is natural to choose $\tilde W=D^{-1}WB$ in~\eqref{eq:w-matrices}, where $D=[WB\mathbf{1}]$ is the diagonal matrix such that $\tilde W$ stochastic.
\end{usecase}
    \begin{usecase}[Social Inertia]\label{ex:inertia}
    A possible explanation for the inclusion of the previous opinion vector $x(t-1)$ is social inertia and status quo bias~\cite{Samuelson1988}, which leads agents to be reluctant to change their beliefs and behaviors. To accommodate this effect, one can choose $\tilde{W} = I$ in~\eqref{eq:w-matrices}.
    \end{usecase}
    \begin{usecase}[Recent Memory Influence]\label{ex:memory}
From the social psychology literature it is known that agents do not immediately forget their neighbors' previous opinion (i.e. recent memory)~\cite{gerrig2013}. Consequently, we may consider $$ W^{(1)}=[\vect 1 -\beta] W \quad , \quad W^{(2)}=[\beta]W, $$
    where $\beta\in\mathbb [0,1]^n$ is a rescaling factor s.t. $1-\beta_i$ is the importance that node $i$ assigns to the updated opinion and  $\beta_i$ reflects their reliance on the previous one.
    \end{usecase}

Finally, we note that the general FJ-MM model can accommodate various other scenarios; for example, $\tilde{W}$ in~\eqref{eq:FJ_HigherOrder} could be a convex combination of the matrices from use cases 2–4, capturing the combined effects of multi-hop influence, inertia, and memory.
The FJ-MM model~\eqref{eq:dynamical system with delay} also applies to cases where memory is caused by communication lag: agents receive messages from others with a one-step lag, while their own opinions remain up-to-date. Formally, in this specific scenario, we can define $W^{(1)} = \text{diag}(W)$ and $W^{(2)} = W - \text{diag}(W)$.

\section{Asymptotic Stability Criterion}\label{sec.stab}

In what follows, we employ the concept of a \emph{comparison} FJ system, which is a standard FJ system~\eqref{eq:dynamical system} defined with a specific stochastic matrix of influence, specifically $W^{(1)}+W^{(2)}$.
\begin{definition}
Given the FJ-MM model~\eqref{eq:dynamical system with delay}, its \textbf{comparison  FJ system} is defined as
\be \label{eqn:HO_withoutdelay} 
 x(t+1)= \bar Ax(t)+(I-\Lambda)s,\;\;\bar A:=\Lambda\left(W^{(1)}+W^{(2)}\right).
\,\ee
\end{definition}

\begin{remark}
The concept of the comparison system extends naturally to the case $L > 2$; in this setting, the comparison model~\eqref{eqn:HO_withoutdelay} generalizes the dynamical model introduced in\cite{Zhang2020_FJwithHigherOrder}. In~\eqref{eqn:HO_withoutdelay}, the matrices $W^{(\ell)}$ can be arbitrary substochastic matrices whose sum is stochastic, and need not correspond to weighted walk matrices as in~\cite{Zhang2020_FJwithHigherOrder}. Unlike the comparison model -- which is a standard FJ model with a modified weight matrix -- the FJ-MM captures influence time lags arising from multi-hop and memory effects.
\end{remark}

It is noteworthy that in the proposed FJ-MM model, $W^{(2)}$ captures the influence of past node opinions, meaning that opinions at time $t+1$ are influenced by those at $t-1$. As a result, the model deviates from the one in \cite{Zhang2020_FJwithHigherOrder}, which can instead be reformulated as an original FJ model with a modified weight matrix.

\subsection{Stability Criteria}

We now focus on the update dynamics introduced in~\eqref{eq:dynamical system with delay}
and examine its asymptotic behavior, establishing conditions for the existence of a unique equilibrium and convergence to it. Leveraging results from positive systems theory~\cite{Liu2009_LinearSystemWithDelay}, we show that the comparison system~\eqref{eqn:HO_withoutdelay} characterizes the asymptotic stability of the FJ-MM and its unique equilibrium.
We start with a technical lemma~\cite{Parsegov_CDC2015,2017_tac_parsegov,ProTempoCaoFriedkin2017} about the Schur stability of the FJ model. 
\begin{lemma}
    Given a diagonal matrix $\vect 0 \leq \Lambda \leq I$ and a row-stochastic matrix $\hat W$, $\rho(\Lambda \hat W)<1$ if and only if, given $\mc G$ the graph associated to matrix $\hat W$, the subset of nodes such that $\lambda_{ii}<1$ is non-empty and globally reachable.
\end{lemma}
\begin{theorem} \label{th:equilibrium_conditions}
The following statements are equivalent:
\begin{itemize}
    \item[(i)] the FJ-MM~\eqref{eq:dynamical system with delay} is exponentially stable;
    \item[(ii)] the comparison FJ system~\eqref{eqn:HO_withoutdelay} is exponentially stable, that is, $\rho(\bar A)<1$;
    \item[(iii)] the subset of nodes $\hat{\mc V}:=\{i \in \mc V : \lambda_{ii}<1\} \subset \mc V$ is non-empty and globally reachable in the graph $\mc G[W^{(1)}+W^{(2)}]$ (that is, the union of $\mc G[W^{(1)}]$ and $\mc G[W^{(2)}]$). This holds, in particular, if $\Lambda<I$.
\end{itemize}
If (i)-(iii) hold, then all solutions of the FJ-MM model~\eqref{eq:dynamical system with delay} and the comparison FJ model~\eqref{eqn:HO_withoutdelay} converge to the point
\be \label{eq:equilibrium} \bar x=\left(I-\bar A \right)^{-1}(I-\Lambda)s,\ee
which serves as the common and unique equilibrium for both systems\footnote{It can be shown~\cite{Friedkin:2015,proskurnikov2017tutorial} that the matrix $\left(I-\bar A \right)^{-1}(I-\Lambda)$, referred in the case of the original FJ model to as the ``control matrix''~\cite{Friedkin:2015} is stochastic, that is, each final opinion is a convex combination of the innate opinions.}. 
\end{theorem}
\begin{proof}
   We first rewrite the dynamical system~\eqref{eq:dynamical system with delay}  as
    \be \label{eq:system_reformulated}
y(t)= \bar A_d y(t-1)+ \bar C.
    \ee 
    where, by definition
    \[\begin{array}{l}
          y(t-1):=\begin{bmatrix}x(t-1) \\ x(t)\end{bmatrix},\quad 
         \bar A_d := \begin{bmatrix}
    0 &I\\ \Lambda W^{(2)}  &\Lambda W^{(1)}
\end{bmatrix},\;
 \bar C := \begin{bmatrix}
    0\\ (I-\Lambda)s
\end{bmatrix}\,.
    \end{array} \]
    
    The system~\eqref{eq:system_reformulated} (equivalent to the FJ-MM) is exponentially stable if and only if the matrix $\bar A_d$ is Schur stable, i.e. $\rho(\bar A_d)<1$. The latter spectral radius, in accordance with the Perron-Frobenius theorem, serves as the maximum real eigenvalue $\lambda\geq 0$ of $\bar A_d$, that is, maximum of $\lambda$ such that  $\bar A_d v = \lambda v$ with some non-zero vector $v = [v_1^{\top} ; v_2^{\top}]^{\top}\ne 0$. 

First notice that since $\bar{A}_d$ is a substochastic matrix, $\rho(\bar{A}_d) \leq 1$.
Explicitly computing now $\bar{A}_d v = \lambda v$, we observe that $v_2=\lambda v_1$, which leads to the following second-order equation:
\be\label{eq:aux}
(\Lambda  W^{(2)}+\Lambda W^{(1)}\lambda)v_1 = \lambda^2 v_1.
\ee 
Recall that $\bar{A}_d$ is Schur stable if and only if 1 is not an eigenvalue of $\bar{A}_d$, or in other words that ~\eqref{eq:aux} has no non-trivial solution $v_1 \neq 0$ when $\lambda = 1$ . This last requirement, referring to \eqref{eq:aux} is equivalent to ask that $\lambda$ is not an eigenvalue for $\bar A$, which from \eqref{eqn:HO_withoutdelay} is equal to $\bar A = \Lambda (W^{(1)}+W^{(2)})$.  Finally, as $\bar{A}$ is also substochastic, this implies that $\rho(\bar{A}) < 1$, and the thesis follows. 
We have thus proven the \textbf{equivalence of (i) and (ii):} the matrices $\bar{A}_d$, which determines the exponential stability of the FJ-MM, and $\bar{A}$, which determines the stability of the comparison FJ model, are either both Schur stable or both have eigenvalue $1$.

The \textbf{equivalence of (ii) and (iii)} is immediate from the criterion for the FJ models stability~\cite{Parsegov_CDC2015,2017_tac_parsegov,ProTempoCaoFriedkin2017}, stating that $\rho(\Lambda \hat W)<1$ for a stochastic matrix $\hat W=W^{(1)}+W^{(2)}$ if and only  {if} the subset of nodes $\hat{\mc V}$ is globally reachable in the graph $\mc G[\hat W]$. Finally, note that a state vector $x$ is an equilibrium of the system~\eqref{eq:dynamical system with delay} or~\eqref{eqn:HO_withoutdelay} if and only if 
$$
x = \bar A x+(I-\Lambda)s,$$
which equation has a unique solution~\eqref{eq:equilibrium} for every $s$ whenever $(I-\bar A)$ is an invertible matrix. 
\end{proof}

In the case where the matrices $W^{(\ell)}$ are decomposed as in~\eqref{eq:w-matrices}, stability can often be tested as follows.
\begin{corollary} \label{cor:suff_condition_rho}
Let $W^{(1)}$, $W^{(2)}$ be defined as in~\eqref{eq:w-matrices}, where $W, \tilde W$ non-negative stochastic matrices and 
$\beta_i \in (0,1)$, for any $i$ in $\mc V$. Then, for the stability of the FJ-MM (equivalently, the comparison FJ model) it suffices that one of the matrices $\Lambda W$ or $\Lambda\tilde W$ is Schur stable.
\end{corollary}
\begin{proof}
The statement is straightforward from Theorem~\ref{th:equilibrium_conditions} model stability, because
the graph $\mc G[W^{(1)}+W^{(2)}]$, obviously, contains both graphs $\mc G[W]$ and $\mc G[\hat W]$, hence, if the nodes from
$\hat{\mc V}$ are globally reachable in one of the graphs, they are also globally reachable in $\mc G[W^{(1)}+W^{(2)}]$.
\end{proof}

\begin{figure}[htb]
    \centering
    \subfloat[{Graph associated to $W$}]{
\includegraphics[width=0.25\linewidth]{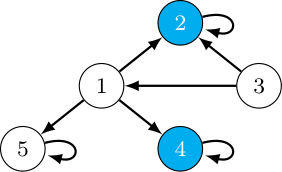} 
\label{fig:example_joint_convergence_a}}\,
\subfloat[{Graph associated to $\tilde W$}]{
\includegraphics[width=0.3\linewidth]{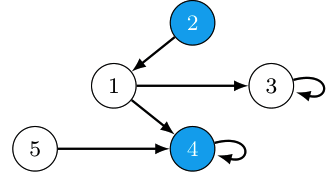} \label{fig:example_joint_convergence_b}}
\caption{
The example of two matrices $W,\tilde W$ with $\rho(\Lambda W)=\rho(\Lambda\tilde W)=1$, resulting in the Schur stable FJ-MM model for $\vect 0<\beta<\vect 1$ (i.e., $\rho(\Lambda([\vect 1 -\beta] W + [\beta] \tilde W))<1$). 
The colored nodes correspond to set $\mathcal{\hat V}=\{i:\lambda_{ii}<1\}$. }
    \label{fig:example_joint convergence}
\end{figure}

It should be noted that Corollary~\ref{cor:suff_condition_rho} provides only a sufficient condition. In fact, it is straightforward to construct an example (see Fig.~\ref{fig:example_joint convergence}) where neither $\Lambda W$ nor $\Lambda\tilde{W}$ is Schur stable, yet the FJ-MM model with matrices~\eqref{eq:w-matrices} is exponentially stable. In other words, the opinion dynamics can be stabilized by introducing memory or multi-hop social influence. A similar delay-induced consensus effect in DeGroot models has been studied in~\cite{ChenLu2017}.

\subsection{Numerical Examples}

Next, we focus on the equilibrium achieved by the FJ-MM model and compare it with that of the original FJ model through numerical simulations. This analysis suggests potential strategies for steering the network’s global equilibrium toward a desired state by appropriately designing (or facilitating) multi-hop interactions.

\begin{figure}[htb]
    \centering
    \includegraphics[width=0.6\linewidth]{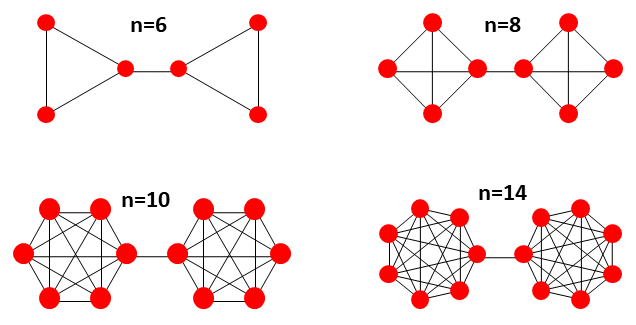}
    \caption{Examples of Barbell Graphs}
    \label{fig:barbell}
\end{figure}

In the following examples we refer to a typical Barbell graph, obtained by connecting two copies of a complete graph by an edge (refer to Fig.~\ref{fig:barbell} for examples). As initial condition, we consider the two complete graphs as two polarized communities, with opinion fixed to 0 and 1, respectively.

All the nodes are assumed to be fully open to social influence (i.e. $\lambda_{ii}=1$), except for those ones for which is differently specified. For the FJ-MM model we will refer in particular to the setting of Use Case \ref{ex:2hop}, that is, $W=W^2$. 

\begin{figure}[htb]
 \centering
\subfloat[original FJ]{
    \begin{tikzpicture}\definecolor{azzurro}{RGB}{0,255,255}
\definecolor{fucsia}{RGB}{255,0,255}
\definecolor{primo}{RGB}{177,78,255}
\definecolor{secondo}{RGB}{78,177,255}

			\node[shape=circle,draw=black,thick,fill=fucsia] (1) at (-.5,0) {\footnotesize1};
			\node[shape=circle,draw=black,thick,fill=fucsia] (2) at (2.5,0) {\footnotesize2};
             \node[shape=circle,draw=black,thick,fill=fucsia] (3) at (1,-0.9) {\footnotesize\color{black}3};

             			\node[shape=circle,draw=black,thick,fill=secondo] (4) at (-.5,-3) {\footnotesize6};
			\node[shape=circle,draw=black,thick,fill=azzurro] (5) at (2.5,-3) {\footnotesize5};
             \node[shape=circle,draw=black,fill=azzurro,thick] (6) at (1,-2.1) {\footnotesize\color{black}4};

 			\node  at (-.6,-.5) {\tiny{$\bar x_1=0$}};
 			\node  at (2.6,-.5) {\tiny{$\bar x_2=0$}}; 			
                \node  at (0.3,-.9) {\tiny{$\bar x_3=0$}};
                \node  at (1.7,-2.1) {\tiny{$\bar x_4=1$}};
             	\node  at (2.6,-2.5) {\tiny{$\bar x_5=1$}}; 	
             	\node  at (-.6,-2.5) {\tiny{$\bar x_6=1$}}; 	
			
    \path [thick,-] (1) edge node[left]{} (3);
    \path [thick,-] (2) edge node[left]{} (3);
    \path [thick,-] (1) edge node[left]{} (2);
    \path [thick,-] (6) edge node[left]{} (3);
    \path [thick,-] (5) edge node[left]{} (4);
    \path [thick,-] (5) edge node[left]{} (6);
    \path [thick,-] (4) edge node[left]{} (6);
		\end{tikzpicture} \label{fig:FJ_withpol}}\subfloat[FJ-MM]{
 \begin{tikzpicture}\definecolor{azzurro}{RGB}{0,255,255}
\definecolor{fucsia}{RGB}{255,0,255}
\definecolor{primo}{RGB}{177,78,255}
\definecolor{secondo}{RGB}{78,177,255}

			\node[shape=circle,draw=black,thick,fill=primo] (1) at (-.5,0) {\footnotesize1};
			\node[shape=circle,draw=black,thick,fill=primo] (2) at (2.5,0) {\footnotesize2};
             \node[shape=circle,draw=black,thick,fill=fucsia] (3) at (1,-0.9) {\footnotesize\color{black}3};

             			\node[shape=circle,draw=black,thick,fill=secondo] (4) at (-.5,-3) {\footnotesize6};
			\node[shape=circle,draw=black,thick,fill=secondo] (5) at (2.5,-3) {\footnotesize5};
             \node[shape=circle,draw=black,fill=azzurro,thick] (6) at (1,-2.1) {\footnotesize\color{black}4};

 			\node  at (-.4,-.5) {\tiny{$\bar x_1\simeq0.31$}};
 			\node  at (2.3,-.5) {\tiny{$\bar x_2\simeq0.31$}}; 			
                \node  at (0.3,-.9) {\tiny{$\bar x_3=0$}};
                \node  at (1.7,-2.1) {\tiny{$\bar x_4=1$}};
             	\node  at (2.3,-2.5) {\tiny{$\bar x_5\simeq0.69$}}; 	
             	\node  at (-.3,-2.5) {\tiny{$\bar x_6\simeq0.69$}}; 	
			
    \path [thick,-] (1) edge node[left]{} (3);
    \path [thick,-] (2) edge node[left]{} (3);
    \path [thick,-] (1) edge node[left]{} (2);
    \path [thick,-] (6) edge node[left]{} (3);
    \path [thick,-] (5) edge node[left]{} (4);
    \path [thick,-] (5) edge node[left]{} (6);
    \path [thick,-] (4) edge node[left]{} (6);
		\end{tikzpicture} \label{fig:FJ_nopol}}
   \caption{Example~\ref{exam.1}: Comparison between the nodes' equilibrium opinions $\bar x_i$ in case of the original FJ and the FJ-MM models. 
   }
    \label{fig:break_pol}
\end{figure}
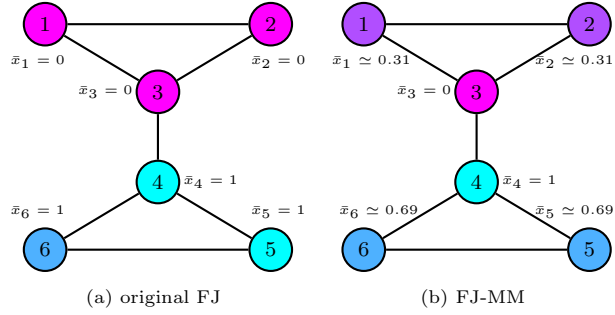

 \begin{example}\label{exam.1}
  Consider a Barbell graph\footnote{Henceforth, for each graph specified below, we construct the matrix $W$ under the assumption that every agent assigns equal weight to all of its neighbors. Consequently, all nonzero entries in each row of $W$ are identical and correspond one-to-one with the arcs emanating from that node.} with 6 nodes. We start from a polarized initial condition, s.t. nodes in \{1,2,3\} are fixed to 0 and \{4,5,6\} to 1. We set $\lambda_{33}=\lambda_{44}=0$, whereas other $\lambda_{ii}=1$, and $[\beta]=0.8I_6$. Under this configuration, the FJ model maintains polarization at equilibrium, whereas the FJ-MM model reduces it, leading to a more balanced equilibrium. In Fig.~\ref{fig:break_pol}, the colormap and labels represent the nodes' equilibrium opinions.      
\end{example}

A formal way to measure polarization is to measure how far we are from the state of complete neutrality. 
One of such measures is the polarization index~\cite{Matakos2017MeasuringAM}, defined as $P=\frac{(\bar x- x^*)^{\top}(\bar x-  x^*)}{n}$, where $\bar x$ is the equilibrium opinions vector and $ x^*= \bar x^{\top} \vect 1/n$ is its mean value. 
In the next example, we compute the polarization index as a function of the rescaling parameter $\beta$ in the network. For simplicity, we assume that $\beta_i=\beta_0$ for all $i\in\mc V$, meaning that agents assign equal total weight to past opinions.
\begin{example}\label{exam.2}
Consider a Barbell graph with 10 nodes. Assume that the initial opinions are polarized as in the previous example, with one clique’s opinions set to $0$ and the other set to $1$. Moreover, the agents at the endpoints of the connecting edge are completely stubborn (i.e., $\lambda_{ii}=0$), while the remaining agents are maximally susceptible (i.e., $\lambda_{ii}=1$) and $[\beta]=\beta_0I$, where $\beta_0$ is changing. Figure~\ref{fig:pol_index_barbell} displays $P$ as a function of $\beta_0$. Our findings indicate that as $\beta_0$ increases -- thereby enhancing the influence of secondary neighbors -- the polarization becomes weaker.
\begin{figure}
    \centering
    \includegraphics[width=0.4\linewidth]{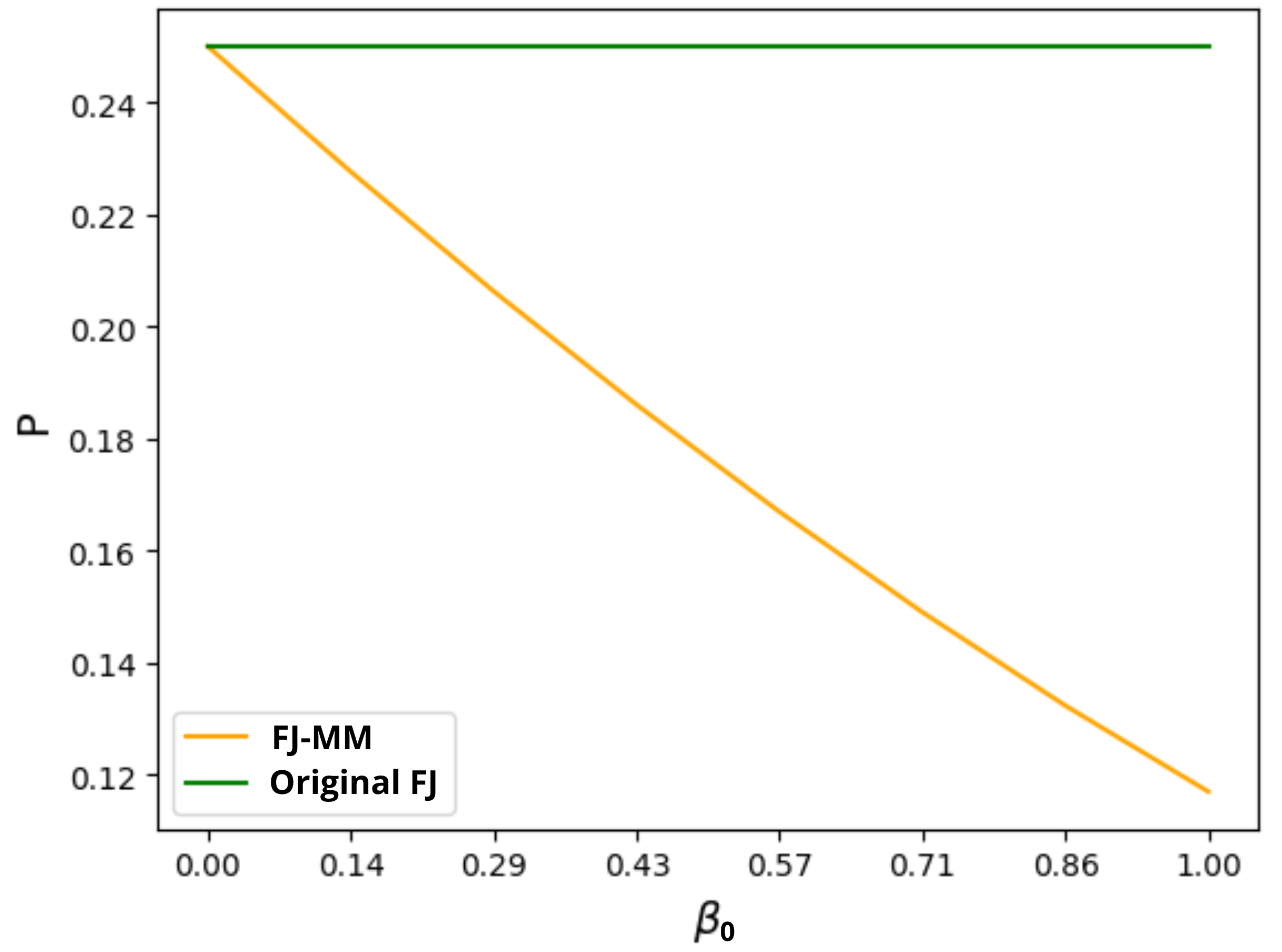}
    \caption{Example~\ref{exam.2}: Polarization Index for a Barbell Graph with 10 nodes.}
    \label{fig:pol_index_barbell}
\end{figure}
\end{example}

Along with polarization, the higher-order neighbors visibly affect the average network's opinion. 
\begin{example}\label{exam.3}   
   Consider a Barbell graph with 16 nodes, using the same polarized initial condition as in Examples 1 and 2. We set $[\beta]=0.8I$, and assign $\lambda_{ii}=0$ to the two nodes at the endpoints of the central edge as well as to two randomly selected nodes (one from each clique). The remaining agents are assigned $\lambda_{ii}=1$. 
   Fig.~\ref{fig:diff_eq} compares the average opinion dynamics for the FJ-MM, the comparison FJ model, and the original FJ model with matrix $W$. It is evident that the FJ-MM model converges more slowly than both the original FJ and the comparison models. We present preliminary results on the convergence rate in the next section. Fig.~\ref{fig:diff_hist} displays the final opinion distributions of individual nodes.
\begin{figure}
 \centering
\subfloat[Network's Average Opinion Dynamics]{
    \includegraphics[width=0.48\linewidth]{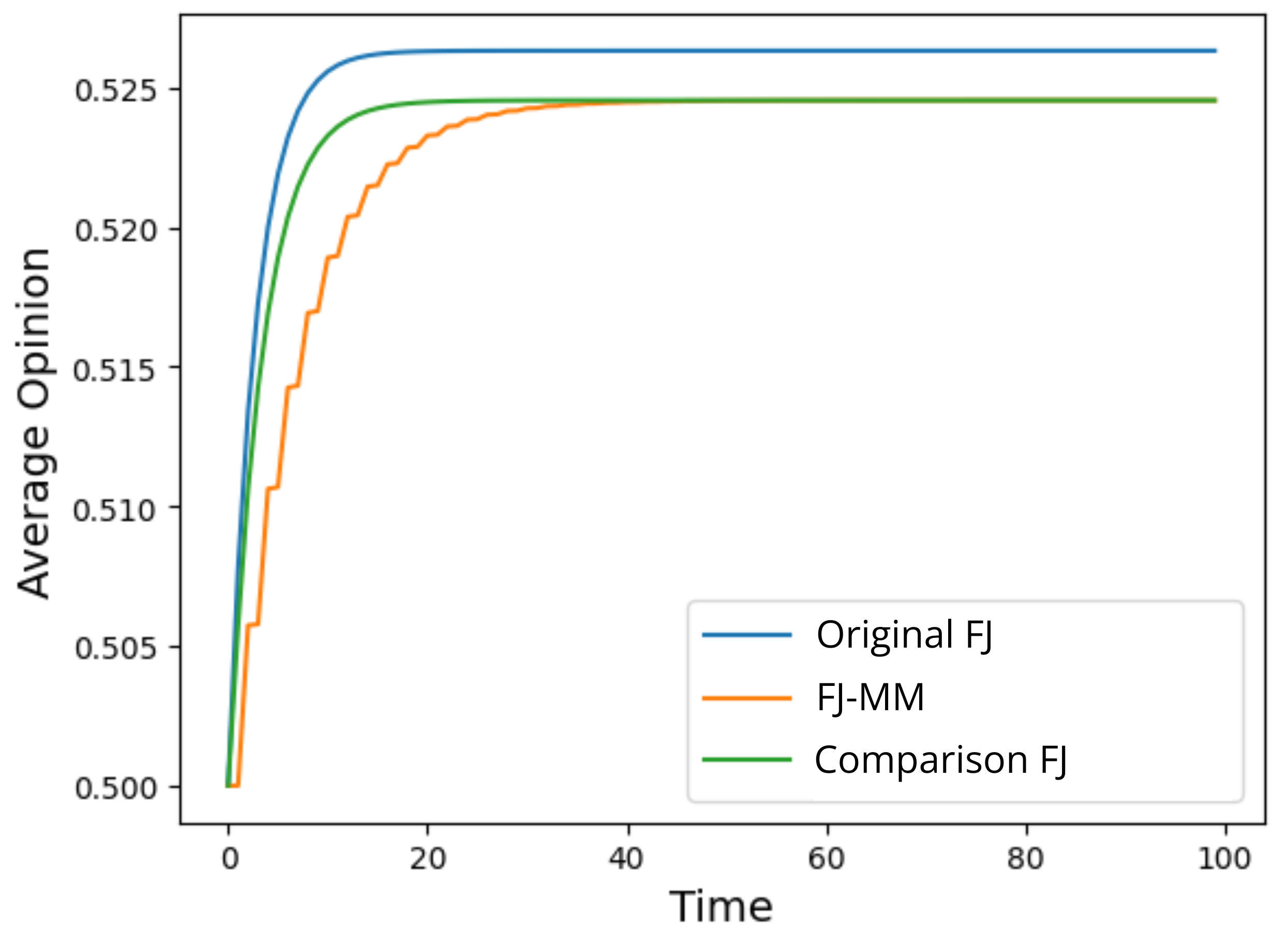}   \label{fig:diff_eq}}\hfill
\subfloat[Nodes' Ultimate Opinions Distribution]{\includegraphics[width=0.45\linewidth]{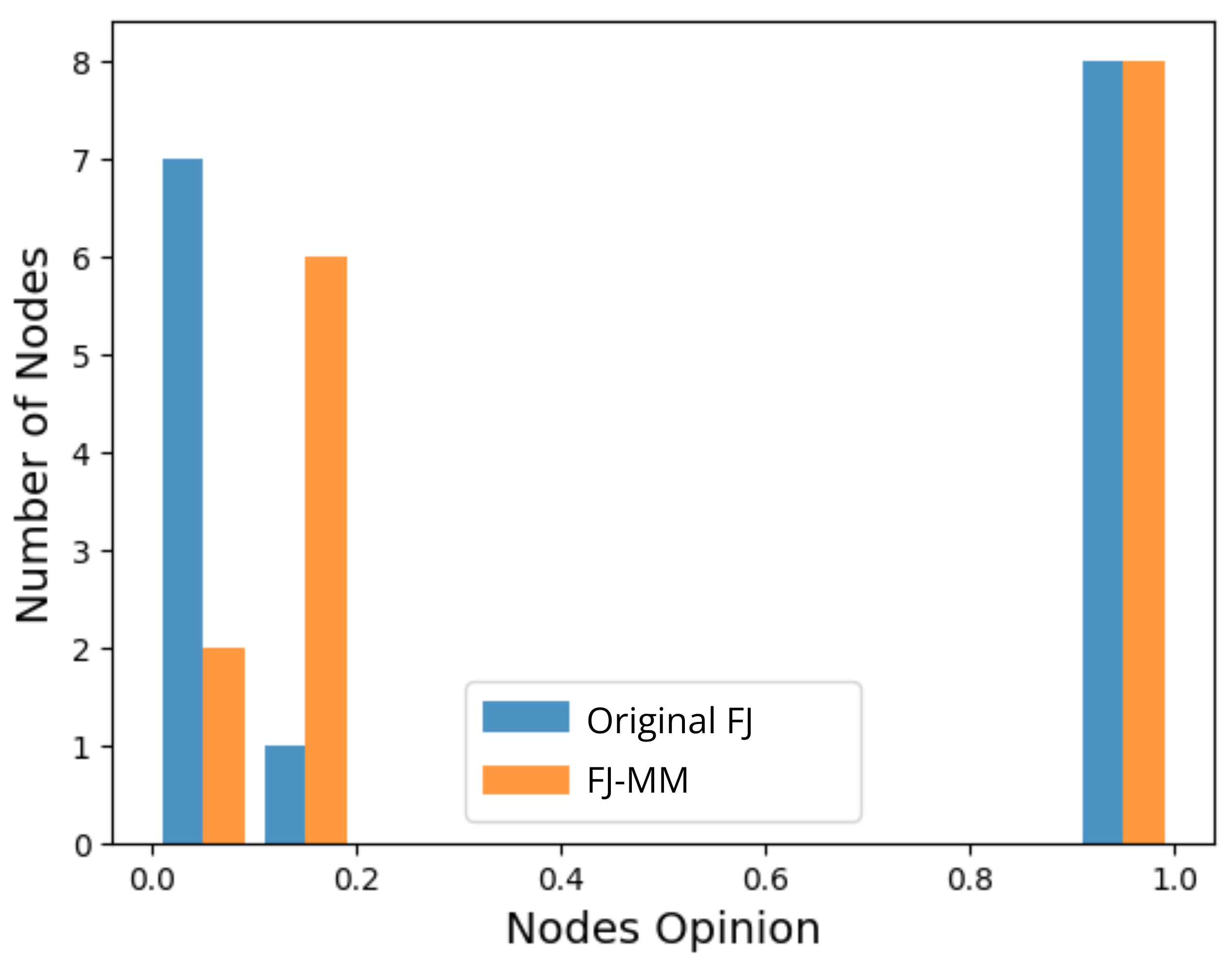}\label{fig:diff_hist}}
   \caption{Example~\ref{exam.3}: Comparison of the averaged opinion's dynamics and the ultimate opinion vectors for the original FJ model~\eqref{eq:dynamical system}, the FJ-MM and the comparison FJ model. 
   }
    \label{fig:different_equilibrium}
\end{figure}
  \end{example}

\section{Convergence Rate}\label{sec.rate}

In the previous section, we have established conditions for the existence of a unique equilibrium for the system introduced in \eqref{eq:dynamical system with delay} and  for convergence to it. Now, we focus on the analysis of the corresponding convergence rate.  Specifically, we analyze the convergence rate for the FJ-MM model~\eqref{eq:system_reformulated} (determined by the spectral radius $\rho(\bar A_d)$) and
compare it with the convergence rates of the comparison FJ model~\eqref{eqn:HO_withoutdelay}. 
In the scenarios of Use Cases 1,3,4, where the FJ-MM is determined by a single stochastic matrix $W$, it is also interesting to compare these convergence rates with the one of the original FJ model~\eqref{eq:dynamical system}.

To simplify the notation, also in this case, we will refer to the FJ-MM model defined in \eqref{eq:dynamical system with delay} using the formulation introduced in \eqref{eq:system_reformulated}, i.e.
$$y(t)=\bar A_d y(t-1) + \bar c \,.$$
Similarly, we will refer to comparison model using \eqref{eqn:HO_withoutdelay}.

First, we will focus on the relation among the convergence rates for the FJ-MM and for the comparison model, showing that the introduced FJ-MM system converges slower to the steady state with respect to the other one.
\begin{proposition} \label{cor:convergence rate delay impact}
Under Assumption \ref{a:2}, the convergence rate of the FJ-MM system in \eqref{eq:dynamical system with delay} does not exceed the one of FJ comparison model in \eqref{eqn:HO_withoutdelay}: $\rho(\bar A_d) \geq \rho(\bar A)$.
\end{proposition}
\begin{proof}
    Consider the eigenvalue-eigenvector system associated to the dynamical system in \eqref{eqn:HO_withoutdelay}.  Given $\rho(\bar A)$ the spectral radius of $\bar A$, it exists $v \in \mathbb{R}^n$ s.t.
    $\bar A v = \rho(\bar A) v, $ with $v$ eigenvector associated to maximum eigenvalue $\rho(\bar A)$. Let us now define $x=[v;v]$. Then, it holds:
    \be \label{eqn:radii_comparison} \bar A_d x =
    \begin{bmatrix}
    0 &I\\ \Lambda W^{(2)}  &\Lambda W^{(1)}
\end{bmatrix} x = \begin{bmatrix}
    v\\ \rho(\bar A) v
\end{bmatrix} \geq \rho(\bar A) x
    \ee
    From Corollary 3.2 in \cite{Marek1990_spectralradii}, it follows that $\rho(\bar A_d) \geq \rho(\bar A)$. 
\end{proof}
The result is in line with what intuitively expected. Indeed, the memory term acts as a rumor that drives the node opinion back to the past, influenced by opinions at previous steps. 
Moreover, focusing on Use Case \ref{ex:memory}, as shown in the following Corollary, the inequality holds also compared to the original FJ.
\begin{corollary} \label{lemma:memory_inequality_rate}
     Let us assume to be under the assumptions of Use Case \ref{ex:memory}, i.e. recent memory influence. The FJ-MM model converges slower than the original FJ model without memory, that is,
    \[
\rho(\bar{A}_d)\geq\rho(\Lambda W)\,.
\]
\end{corollary}
\begin{proof}
     Proposition~\ref{cor:convergence rate delay impact}, since $W^{(1)}+W^{(2)}=W$, implies that $\bar A=\Lambda W$ and $\rho(\bar A)=\rho(\Lambda W)$.
\end{proof}
\subsection{A Special Case: Homogeneous Susceptibility}

It is interesting to explicitly compare the convergence rates of the FJ-MM model with matrices~\eqref{eq:w-matrices} and its corresponding comparison FJ model. This comparison is relatively straightforward when all agents share the same level of susceptibility to social influence, $\lambda_{ii}$, and assign identical total weights, $\beta_i>0$, to past opinions: 
\begin{assumption}\label{asm.homogen}
    Assume that $\Lambda = \sigma I$ with $\sigma \in \mathbb{R}$ with\footnote{The case of $\Lambda=0$ is, obviously, degenerate, as all agents are totally stubborn and the dynamics terminate in a single step. The case $\Lambda=I$ corresponds to the DeGroot model, where there is no asymptotic stability.} $\sigma\in(0,1)$. Moreover, assume that all elements of the vector $\beta$ are equal $\beta_i=\beta_0\in(0,1)$ for every $i \in \mc V$. 
\end{assumption}

The spectral radius of the comparison FJ model with such a susceptibility matrix is straightworward to find.
\begin{lemma} \label{lemma:homogeneous_eq_rho}
If $\Lambda=\sigma I$, then for every stochastic matrix $\hat W$ one has $\rho(\Lambda \hat W)=\sigma$, in particular,
$\rho(\bar{A})=\sigma$. Hence, the convergence rate of the FJ-MM admits the following lower bound
\[
\rho(\bar{A}_d)\geq \rho(\bar{A})=\sigma.
  \]
\end{lemma}
\begin{proof}
    The first statement is straightforward since $\rho(\sigma\hat W)=\sigma$. The second statement follows from  Corollary~\ref{cor:convergence rate delay impact}.
\end{proof}
%
By virtue this lemma, it can be easily shown that the inequality in Proposition \ref{cor:convergence rate delay impact} holds strictly in several cases.
\begin{proposition}\label{prop:convergence_linear_comb}
   Consider the FJ-MM with matrices~\eqref{eq:w-matrices}, where $W$ is some stochastic matrix and
   $\tilde W$ characterizes the joint impact of social inertia (as in Use Case \ref{ex:inertia}) and recent memory influence (as in Use Case \ref{ex:memory}). Specifically,
\[
\tilde W = \alpha_1 W + \alpha_2 I,\quad \alpha_1,\alpha_2\geq 0,\quad \alpha_1+\alpha_2 =1.
\]
Suppose also that Assumption~\ref{asm.homogen} holds. Then the FJ-MM model converges strictly slower than the comparison FJ model and than the original FJ model with matrix $W$. More precisely,
\begin{equation}\label{eq:spec1-memory}
\rho(\bar{A}_d)=\frac{\sigma(1-\beta_0)+\sqrt{\sigma (1-\beta_0)^2+4\beta_0}}{2}>\rho(\bar A)=\rho(\Lambda W)=\sigma.
\end{equation}
\end{proposition}
\begin{proof}
We first compute the spectral radius for the FJ-MM model. 
Recalling that the eigenvalues of $\bar A_d$ are such scalars $\lambda$ such that $\bar A_d v = \lambda v$, with some vector $v = [v_1^{\top} ; v_2^{\top}]^{\top}$, then observing that $v_2 = \lambda v_1$, one proves that
\be\label{eq:aux_inertia}
(\Lambda \beta W+\Lambda (1-\beta_0)(\alpha_1 W+\alpha_2I)\lambda)v_1 = \lambda^2 v_1.
\ee
Substituting $\Lambda=\sigma I$ and using the Perron-Frobenius theorem, the spectral radius $\rho (\bar A_d)$ is the maximum of numbers $\lambda\geq 0$ such that
\begin{equation}\label{eq.aux1}
0=\det(\lambda^2 I-\lambda\sigma (1-\beta_0)W-\sigma \beta_0 (\alpha_1 W+\alpha_2I))=
\det((\lambda^2-\sigma\beta_0\alpha_2)I-[\lambda\sigma (1-\beta_0)+\sigma \beta_0 \alpha_1]),
\end{equation}
Notice that $\lambda\sigma(1-\beta_0) + \sigma\beta_0\alpha_1 > 0$, because if it were not, then $\lambda = 0$ and $\alpha_1 = 0$, which would imply $\lambda^2 - \sigma\beta_0\alpha_2 = -\sigma\beta_0 < 0$. Consequently, $\lambda$ could not be a root of equation~\eqref{eq.aux1}. Hence, the number
\begin{equation}\label{eq.aux1-mu}
\mu:=\frac{\lambda^2-\sigma\beta_0 \alpha_2}{\lambda\sigma (1-\beta_0)+\sigma \beta_0 \alpha_1}\in\mathbb{R}
\end{equation}
$\mu$ appears as an eigenvalue of $W$. Conversely, if $\mu$ is an eigenvalue of $W$ and satisfies~\eqref{eq.aux1-mu}, then $\lambda$ is an eigenvalue of $\bar{A}_d$. By treating~\eqref{eq.aux1-mu} as a quadratic equation in $\lambda$ and selecting its maximal (nonnegative) root, one can show that the spectral radius $\rho(\bar{A}_d)$ is given by
\[
\rho(\bar{A}_d)=\max_{\mu}\frac{\sigma(1-\beta_0)}{2}\left( \mu + \sqrt{\mu^2+\frac{4\beta_0 (\alpha_1\mu + \alpha_2)}{\sigma (1-\beta_0)^2}} \right),
\]
which maximum is taken over all real $\mu$ eigenvalues of $W$ and, obviously, is achieved at $\mu=1$.

Recalling that $\alpha_1+\alpha_2=1>\sigma$, one finally notices that
\[
\begin{aligned}
\rho(\bar{A}_d)=\frac{\sigma(1-\beta_0)}{2}\left( 1 + \sqrt{1+\frac{4\beta_0 (\alpha_1 + \alpha_2)}{\sigma (1-\beta_0)^2}} \right)>\frac{\sigma(1-\beta_0)}{2}\left( 1 + \sqrt{1+\frac{4\beta_0}{(1-\beta_0)^2}} \right)=\sigma,
\end{aligned}
\]
which proves~\eqref{eq:spec1-memory} and finishes the proof of Proposition~\ref{prop:convergence_linear_comb}.
\end{proof}

It is interesting to note that under Assumption~\ref{asm.homogen}, the spectral radius of $\bar{A}_d$ is independent of the coefficients $\alpha_1$ and $\alpha_2$, e.g., the FJ-MM model exhibits the same convergence rate in Use Case \ref{ex:inertia} as in Use Case \ref{ex:memory}.

\begin{figure}[htb]
 \centering
\includegraphics[width=0.48\linewidth]{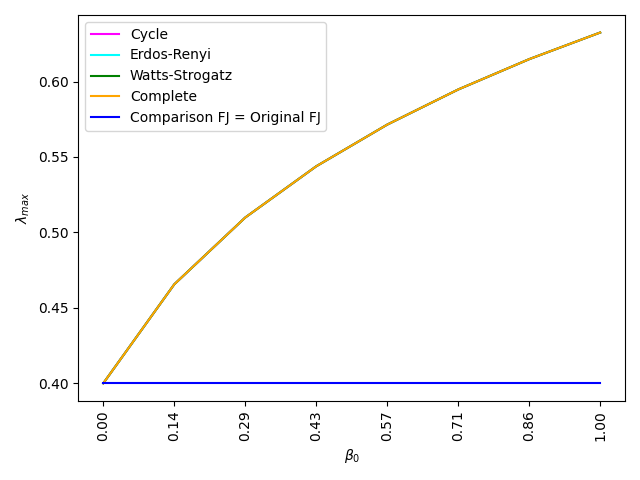}   
\caption{Comparison of spectral radius for FJ-MM vs comparison model vs original FJ in Use Case \ref{ex:inertia} for the homogeneous case. Spectral radius of FJ-MM computed for different network topologies.}
    \label{fig:hom_case}
\end{figure}

The following numerical example illustrates that, for $\Lambda=\sigma I$ with $\sigma$ being fixed and $[\beta]=\beta_0 I$, the spectral radius in Use Case \ref{ex:inertia} depends only on $\beta_0$ but not on the graph, as guaranteed by Proposition~\ref{prop:convergence_linear_comb}.
\begin{example}\label{example.homogen}
     Fig.~\ref{fig:hom_case} illustrates the numerical simulations, in which the convergence rate (maximal eigenvalue) of the FJ-MM is computed for several graphs as a function of $\beta_0$, assuming that $\Lambda=0.6 I$ and $[\beta]=\beta_0 I$. Technically, the example shows the results for an undirected cycle graph with $N=20$ nodes, an Erdos-Renyi graph with $N=150$ nodes and probability of edge creation $0.4$, a Watts-Strogatz graph with $N = 200$ nodes, degree equal to $0.6 N$ and rewiring probability $0.7$, and a complete graph with $N=50$ nodes.
     
    It can be observed that the convergence rate is, first, independent of the graph structure and, second, strictly greater than that of the comparison FJ model (and of the original FJ model with matrix $W$), which equals $\sigma=0.6$.
\end{example}

Using an analogous proof strategy, we can prove that the convergence result holds also in Use Case~\ref{ex:2hop}. 
\begin{proposition}\label{prop:convergence_2hop}
 Consider the FJ-MM with matrices~\eqref{eq:w-matrices}, where $W$ is some stochastic matrix and
   $\tilde W=W^2$ (as in Use Case \ref{ex:2hop}). Suppose also that Assumption~\ref{asm.homogen} holds. Then, the FJ-MM model converges slower than the comparison FJ model (and than the original FJ model with matrix $W$). Namely,
    \[
\rho(\bar{A}_d)=\frac{\sigma (1-\beta_0)+\sqrt{\sigma^2(1-\beta_0)^2+4\sigma \beta_0}}{2}>\rho(\bar A)=\rho(\Lambda W)=\sigma.
\]
\end{proposition}
\begin{proof}
Repeating the argument from the proof of Proposition~\ref{prop:convergence_linear_comb}, we have that $\lambda$ is an eigenvalue of $\bar{A}_d$ if and only if
\be\label{eq:aux_case1}
(\Lambda \beta_0 W+\Lambda (1-\beta_0)W^2\lambda)v_1 = \lambda^2 v_1.
\ee
for some non-zero vector $v_1$. Substituting $\Lambda=\sigma I$, this is equivalent to the relation
 \[
\det(\lambda^2 I -\lambda \sigma(1-\beta_0)W-\sigma\beta_0 W^2)=0.
\]
Let $\mu_1, \ldots, \mu_n$ denote the eigenvalues of $W$. For any analytic function $f:\mathbb{C}\to\mathbb{C}$ -- in particular, for the polynomial
$
f(z)=\lambda \sigma(1-\beta_0)z+\sigma\beta_0 z^2,
$ -- the eigenvalues of $f(W)$ are given by $f(\mu_i)$. Hence, for some $\mu=\mu_i$, one has
\begin{equation}\label{eqn:2hop_condition}
\lambda^2 = \lambda \sigma(1-\beta_0)\mu + \sigma\beta_0 \mu^2.
\end{equation}
Conversely, if $\lambda$ is a solution to~\eqref{eqn:2hop_condition} with some $\mu\in\{\mu_1,\ldots,\mu_n\}$, then
$\lambda$ is an eigenvalue of $\bar A_d$. 

Considering~\eqref{eqn:2hop_condition} as a quadratic equation in $\mu$, one notices that, for $\lambda$ being real, one has
$(-\lambda^2)\leq 0$, and hence~\eqref{eqn:2hop_condition} can only be satisfied for \emph{real} $\mu$. Resolving~\eqref{eqn:2hop_condition} with respect to $\lambda$ for $\mu$ being real, one obtains
\[
\lambda = \frac{\sigma (1-\beta_0)\mu \pm |\mu|\sqrt{\sigma^2(1-\beta_0)^2+4\sigma\beta_0}}{2}.
\]
Hence, $\rho(\bar{A}_d)$, the maximal real eigenvalue of $\bar{A}_d$, is equal to the maximum of these expressions over all real eigenvalues of $W$. Clearly, this maximum is attained when $\mu = 1$ and the larger of the two roots is selected, i.e.,
\[
\rho(\bar A_d)=\frac{\sigma (1-\beta_0)+\sqrt{\sigma^2(1-\beta_0)^2+4\sigma\beta_0}}{2}>
\frac{\sigma (1-\beta_0)+\sqrt{\sigma^2(1-\beta_0)^2+4\sigma^2\beta_0}}{2}=\sigma,
\]
which finishes the proof.
\end{proof}

\subsection{Numerical Analysis}

In general, the convergence rates of the FJ-MM model--as well as its corresponding comparison FJ system--depend significantly not only on the vector $\beta$, but also on the graphs corresponding to $W$ and $\tilde{W}$ and the structure of the susceptibility matrix $\Lambda$. Moreover, their dependence on $\beta$ is generally non-monotonic. Since providing an analytical description of these dependencies is nontrivial, we only demonstrate several numerical experiments.
In these experiments, we use one of the graps, described in Example~\ref{example.homogen}, however, the diagonal entries of $\Lambda$ are now heterogeneous: nodes belonging to a randomly chosen subset $\hat{\mc V}$ are assigned $\lambda_{ii}=0$ (indicating complete stubbornness), while the remaining agents are assigned $\lambda_{ii}=1$.

First, we focus on the dependence of the convergence rate on the network choice under the hypothesis of Use Case~\ref{ex:inertia} (i.e., $\tilde W=I$). Our initial experiment demonstrates that when setting $[\beta]=\beta_0 I$ in~\eqref{eq:w-matrices}, the dependence of the convergence rates on $\beta_0$ differs substantially from the homogeneous $\lambda_{ii}$ case (Proposition 3). In particular, the gap between the spectral radii of the FJ-MM and its comparison models is non-monotonic.
\begin{example} We compare the convergence rates of the FJ-MM model and the comparison FJ model for different network choices under the assumption that $\tilde W=I$ (Social Inertia). We adopt the same graphs as in Example~\ref{example.homogen}. The 
    set of randomly chosen stubborn nodes has cardinality $|\hat{\mc V}|=0.2|\mc V|$). It can be observed (Fig.~\ref{fig:network dependence}) that the gap between the spectral radii of the FJ-MM and the comparison FJ model is no longer a monotonic function of $\beta_0$. This outcome is expected since, when $\beta_0=1$, both models lose asymptotic stability, yielding $\rho(\bar{A}_d)=\rho(\bar{A})=1$. Unlike the situation in Example~\ref{example.homogen}, the gap visibly depends on the network's topology.
    \begin{figure}
 \centering
\includegraphics[width=0.4\linewidth]{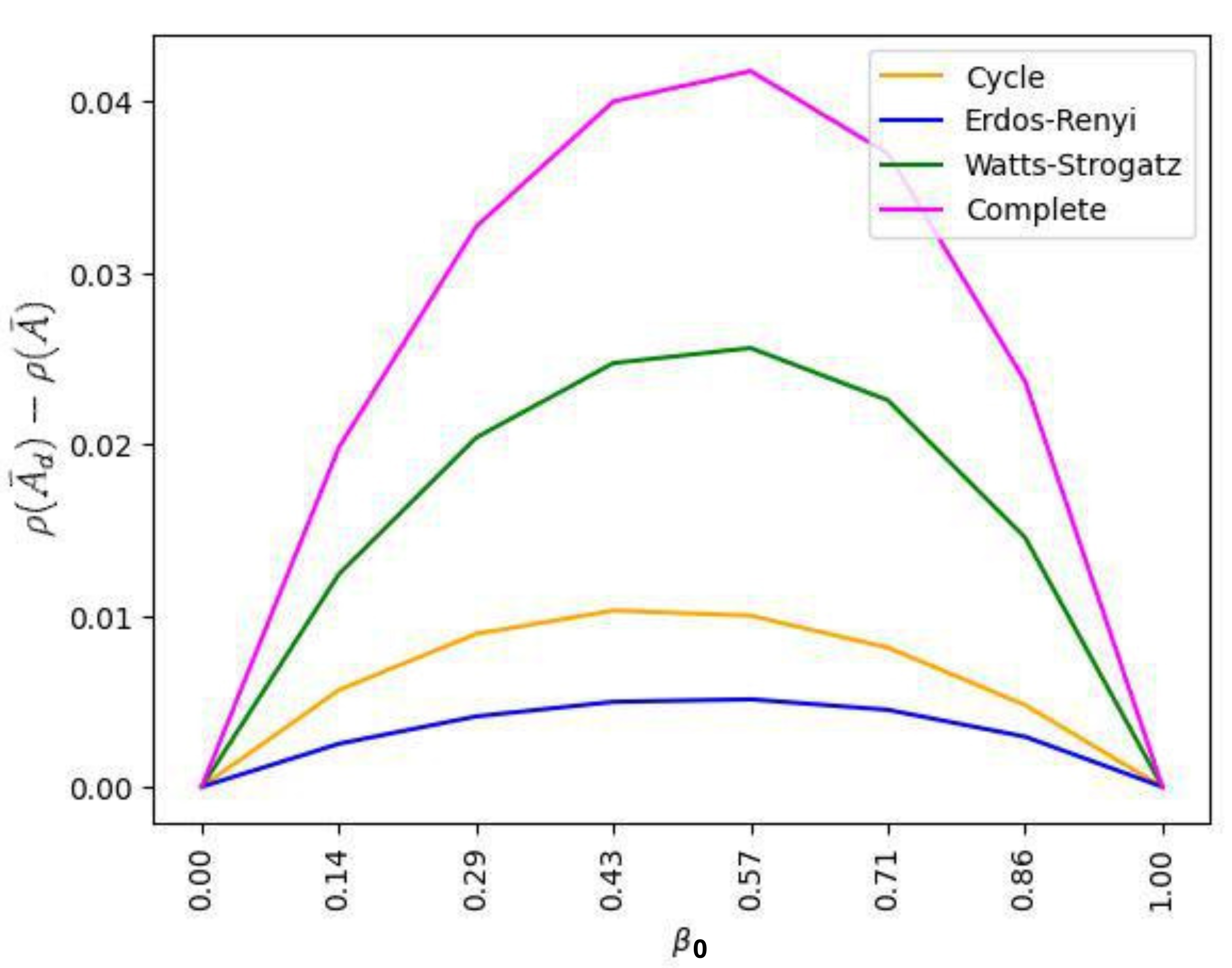}   
\caption{Dependence of convergence rate  on network choices in Use Case \ref{ex:inertia} (FJ-MM vs comparison FJ)}
    \label{fig:network dependence}
\end{figure}
\end{example}



In the case of secondary neighbors (Use Case~1), the relationship between the spectral radii and $\beta_0$ becomes even more complex and substantially depends on $\Lambda$. Specifically, while the spectral radius of the comparison model exhibits a pronounced minimum at some $\beta_0=\beta_*\in (0,1)$, the spectral radius of the FJ-MM model increases monotonically to $1$ as $\beta_0\to 1$. Furthermore, the gap between the two models widens as the number of stubborn agents, $|\hat{\mc V}|$, increases. This behavior is illustrated in our final example.
\begin{example}
[Influence of cardinality of $\hat{\mc V}$]
Consider a Watts-Strogatz graph randomly generated, with $N = 200$ nodes, degree equal to $0.6 N$ and rewiring probability $0.7$, under the assumption that $\tilde W = W^2$ (Use Case~\ref{ex:2hop}). 
We compare two cases: $15\%$ of individuals are stubborn vs. $50\%$ of stubborn agents and show the spectral radii of
the FJ-MM, the associated comparison model and the original FJ model with  matrix $W$ (corresponding to $\beta_0=0$).    
\end{example}
\begin{figure}[htb]
 \centering
\subfloat[$|\hat{\mc V}|=0.15|\mc V|$]{
    \includegraphics[width=0.45\linewidth]{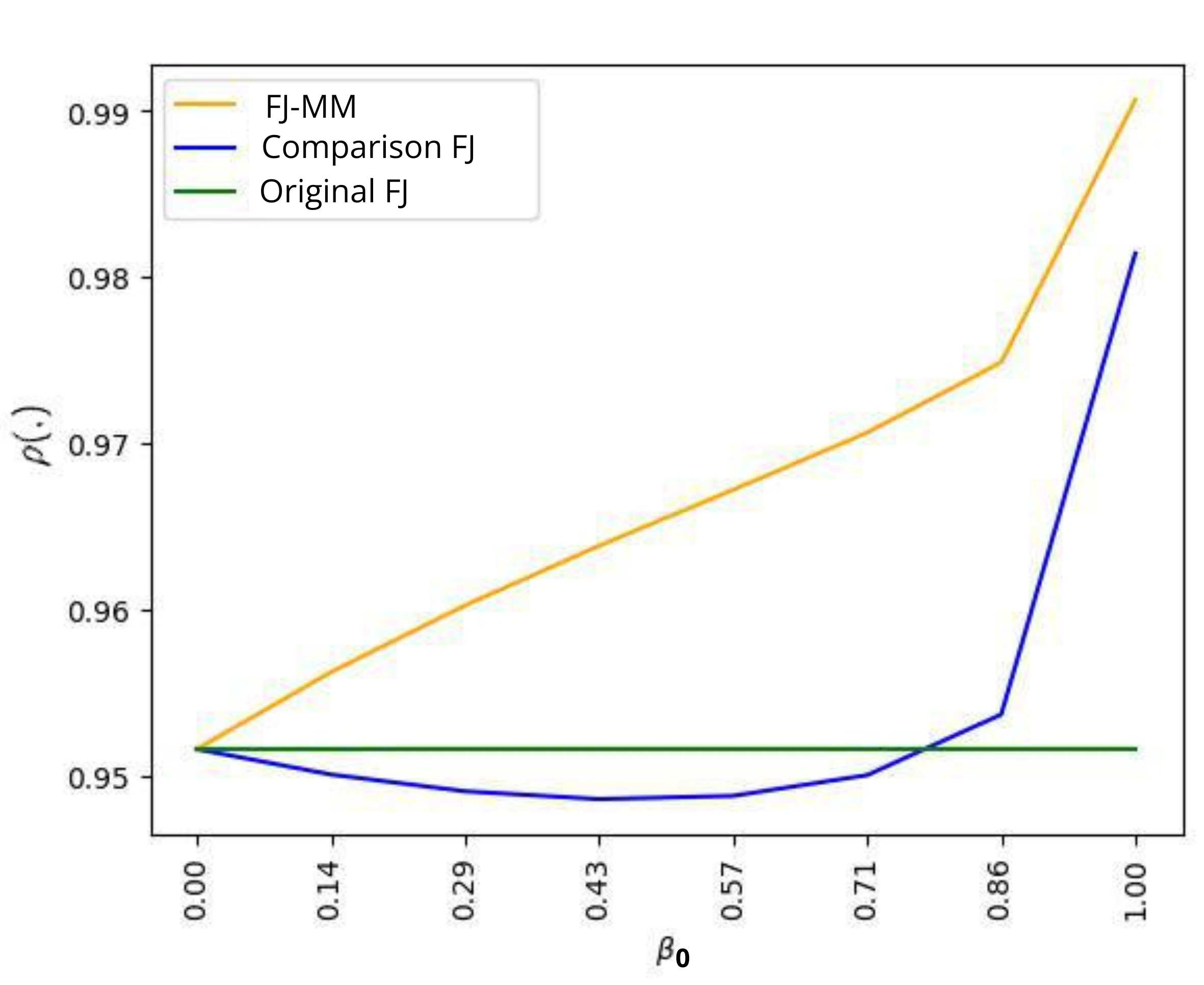}   \label{fig:WS_ex1}}
\subfloat[$|\hat{\mc V}|=0.5|\mc V|$]{\includegraphics[width=0.45\linewidth]{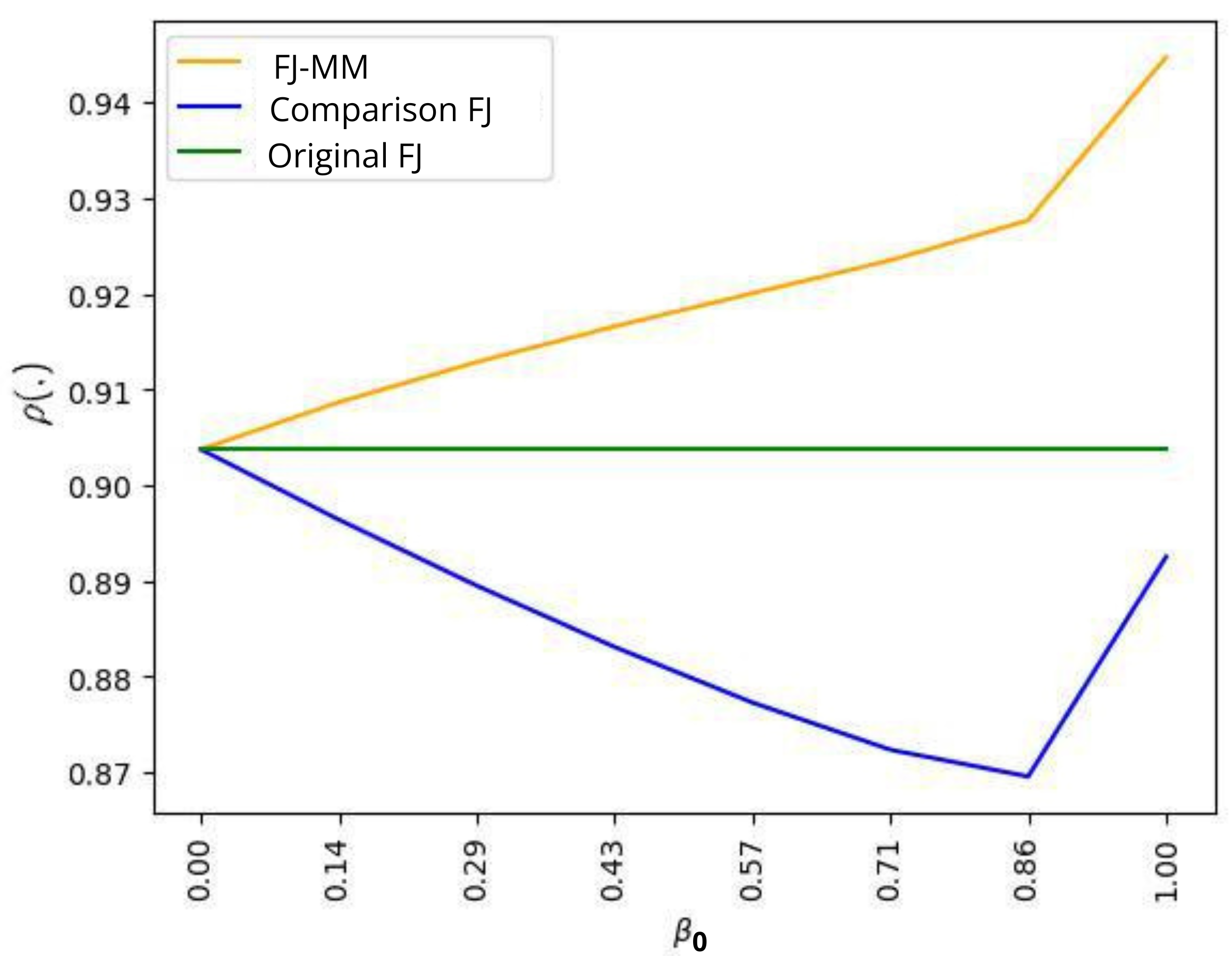}\label{fig:WS_ex2}}
   \caption{Comparison between the maximum eigenvalue in original FJ, FJ-MM model and comparison model for different choices of the cardinality of the set $\hat{\mc V}$ under Use Case \ref{ex:2hop}. 
   }
    \label{fig:cardinality dependence}
\end{figure}

\section{Conclusion}\label{sec.concl}

In this paper, we propose a generalization of the Friedkin–Johnsen (FJ) model, termed FJ-MM, which integrates memory effects and higher-order (multi-hop) neighbor influences to account for both current and past opinions across direct and secondary connections. Our analysis shows that while the convergence properties of the FJ-MM model reduce to those of a comparison model--namely, the standard FJ model with a modified influence matrix--the convergence rate is significantly affected by the incorporation of past opinions, as demonstrated by preliminary eigenvalue analysis and numerical simulations on random graphs. Also, our numerical experiments reveal that memory and multi-hop influence reshape the opinion landscape by reducing polarization in the final opinion profile.

The findings presented in this paper suggest several promising directions for future research, particularly in exploring both the steady-state and transient properties of the FJ-MM dynamics. It is well established that the FJ model naturally gives rise to a centrality measure on influence networks, with PageRank emerging as a special case~\cite{proskurnikov2017tutorial,Friedkin:2015}. A corresponding centrality measure can be defined for the FJ-MM model, naturally prompting the question of how memory and higher-order neighbors influence a node's centrality.
Even for classical FJ models, the relationship between the convergence rate (i.e., the spectral radius) and the properties of influence networks has not been fully explored (some relevant results can be found, e.g., in~\cite{GHADERI20143209,ProTempoCaoFriedkin2017}) -- a challenge that becomes even more pronounced for the FJ-MM model.
While the limitation $L=2$ (capturing only the influence of neighbors-of-neighbors, or one-step memory) can be easily relaxed by considering longer walks, it is plausible that the effective depth of memory is both time-varying and potentially random. For example, social media platforms like Facebook can randomly retrieve events or posts from several months or years ago.
This type of randomness in social interactions is notably distinct from that observed in randomized gossip-based models~\cite{2017_tac_parsegov} and appears to be underexplored.

\bibliographystyle{IEEEtran}
\bibliography{ref}
\end{document}